\documentclass[pre,10pt,twocolumn]{revtex4}%
\usepackage{amsfonts}
\usepackage{amsmath}
\usepackage{amssymb}
\usepackage{graphicx}%
\newtheorem{theorem}{Theorem}

\newenvironment{proof}[1][Proof]{\noindent\textbf{#1.} }{\ \rule{0.5em}{0.5em}}
\begin{document}
\title{Logic circuits from zero forcing }
\author{Daniel Burgarth}
\email{daniel@burgarth.de}
\affiliation{Institute of Mathematics and Physics, Aberystwyth University, SY23 3BZ
Aberystwyth, United Kingdom}
\author{Vittorio Giovannetti}
\email{v.giovannetti@sns.it}
\affiliation{NEST, Scuola Normale Superiore and Istituto Nanoscienze-CNR, Piazza dei
Cavalieri 7, I-56126 Pisa, Italy}
\author{Leslie Hogben}
\email{lhogben@iastate.edu; hogben@aimath.org}
\affiliation{Department of Mathematics, Iowa State University, Ames, IA 50011, USA, and
American Institute of Mathematics, 360 Portage Ave, Palo Alto, CA 94306, USA}
\author{Simone Severini}
\email{simoseve@gmail.com}
\affiliation{Department of Computer Science, and Department of Physics \& Astronomy,
University College London, WC1E 6BT London, United Kingdom}
\author{Michael Young}
\email{myoung@iastate.edu}
\affiliation{Department of Mathematics, Iowa State University, Ames, IA 50011, USA}

\begin{abstract}
We design logic circuits based on the notion of zero forcing on graphs; each
gate of the circuits is a gadget in which zero forcing is performed. We show
that such circuits can evaluate every monotone Boolean function. By using two
vertices to encode each logical bit, we obtain universal computation. We also
highlight a phenomenon of \textquotedblleft back forcing\textquotedblright\ as
a property of each function. Such a phenomenon occurs in a circuit when the
input of gates which have been already used at a given time step is further
modified by a computation actually performed at a later stage. Finally, we
show that zero forcing can be also used to implement reversible computation.
The model introduced here provides a potentially new tool in the analysis of
Boolean functions, with particular attention to monotonicity.

\end{abstract}
\maketitle

\section{Introduction}

We order the two elements of a set $\Sigma=\{0,1\}$ such that $0<1$. This
extends to a partial ordering on the set $\Sigma^{n}=\{0,1\}^{n}$ by comparing
words coordinate-wise. Let $x=x_{1}...x_{n}$ and $y=y_{1}...y_{n}$. Here,
$x\succeq y$ means that $x_{i}\geq y_{i}$, for every $i=1,...,n$. A Boolean
function $f:\Sigma^{n}\longrightarrow\Sigma$ is \emph{monotone} when $f\left(
x\right)  \geq f\left(  y\right)  $ if $x\succeq y$, for every $x,y\in
\Sigma^{n}$.

Monotone Boolean functions have an important role for proving lower bounds of
circuit complexity (see, \emph{e.g.}, Leeuwen \cite{l90}, Chapter 14.4). Any
function obtained by composition of monotone Boolean functions is itself
monotone. Examples of monotone Boolean functions are the conjuction
{\scriptsize AND} and the disjunction {\scriptsize OR}. Indeed, every monotone
Boolean function can be realized by {\scriptsize AND} and {\scriptsize OR
}operations (but without {\scriptsize NOT}). Boolean functions are important
in applications, for example, in the implementation of a class of non-linear
digital filters called stack filters \cite{as}. Important methods for
obtaining non-trivial bounds on specific monotone Boolean functions have been
studied (see, \emph{e.g.}, \cite{al}).

The concept of \emph{zero forcing} on graphs is a recent idea that is part of
a program studying minimum ranks of matrices with specific combinatorial
constraints. Zero forcing has been also called graph infection and graph
propagation \cite{bg, s}. Notice that, in the context described here, the term
\textquotedblleft zero forcing\textquotedblright\ seems to be unfortunate,
because we are forcing ones, not zeros. However, we keep the term given that
this is now the most commonly used in the literature. In order to define zero
forcing, we first need to define a \emph{color-change rule}: if $G=(V,E)$ is a
graph with each vertex colored either white or black, $u$ is a black vertex of
$G$, and exactly one neighbor $v$ of $u$ is white, then change the color of
$v$ to black. Given a coloring of $G$, the \emph{final coloring} is the result
of applying the color-change rule until no more changes are possible. A
\emph{zero forcing set} for $G$ is a set $Z\subseteq V\left(  G\right)  $ such
that if the elements of $Z$ are initially colored black and the elements of
$V(G)\backslash Z$ are colored white, the final coloring of $G$ is all black.

Zero forcing is related to certain minimum rank/maximum nullity problems of
matrices associated to graphs (see \cite{min}) and to the controllability of
quantum spin systems \cite{bg, bs}. Minimimizing the size of zero forcing sets
is a difficult combinatorial optimization problem \cite{aa}.

The remainder of this paper is organized as follows. In Section 2, we prove
that zero forcing on graphs realizes all monotone Boolean functions, and
highlight some simple related facts. The connection between zero forcing and
circuits is obtained by associating a graph to each logic gate. We will show
that the functions {\scriptsize AND} and {\scriptsize OR} are indeed easily
realized by two different gadgets with a few vertices. This is not the first
work observing that monotone Boolean functions can be realized in a
combinatorial setting. For example, Demaine \emph{et al. }\cite{de} have used
the movements of a collections of simple interlocked polygons.

In Section 3, we describe the phenomenon of \emph{back forcing} in the
circuit. The phenomenon occurs when the color-change rule acts to modify the
color of a vertex which has been already used during the computation. In some
cases, back forcing implies that information about the output of a Boolean
circuit can be read not just by looking at the color of a \emph{target} vertex
corresponding to the final output of the process, but at the color of the
vertices in certain intermediate or initial gadgets. The idea opens a simple
but intriguing scenario consisting of many parties that perform computation in
a distributed way: each party holds a subset of the gates and it is able to
read certain information about the input of other parties, since the color of
its gates may have been modified by back forcing. Back forcing can be avoided
by including some extra gadget acting as a filter.

In Section 4, we show that zero forcing
becomes \emph{universal}, \emph{i.e.}, it can realize any Boolean function, if
we apply a proper encoding. Specifically the \emph{dual rail encoding}, where
two vertices are assigned to each logical bit, is a method to construct the
{\scriptsize NOT }gate and therefore to obtain universal computation.
Conclusions are in Section~5.

\section{Main result}

Our main result is easy to prove:

\begin{theorem}
Zero forcing realizes all monotone Boolean functions.
\end{theorem}

\begin{proof}
It is sufficient to show that zero forcing realizes the functions
{\scriptsize AND} and {\scriptsize OR}.

\emph{Claim 1.} The gate {\scriptsize AND} is realized by the gadget
$G_{\text{{\scriptsize AND}}}$ with vertices $\{1,2,3\}$ and edges
$\{\{1,2\},\{1,3\},\{2,3\}\}$, where $1$ and $2$ are the input vertices and
$3$ is the output vertex, containing the result and being able to propagate
the color. All vertices are initially colored white. An illustration of the
gadget $G_{\text{{\scriptsize AND}}}$ is below:%

\begin{figure}
[h]
\begin{center}
\includegraphics[
height=0.7185in,
width=1.499in
]%
{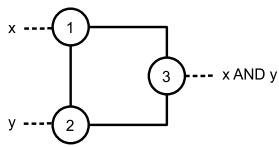}%
\caption{The gate for the function {\scriptsize AND}. }%
\label{and}%
\end{center}
\end{figure}

\emph{Proof of Claim 1. }If no action is taken then the final coloring of the
gadget is white. If we color vertex $1$ black then the final coloring is all
white but for vertex $1$. The same holds for vertex $2$. However, if we color
vertex $1$ and vertex $2$ black then the color-change rule implies that vertex
$3$ is black at step $2$. In fact, $\{1,2\}$ is a zero forcing set for
$G_{\text{{\scriptsize AND}}}$.

\emph{Claim 2. }The gate {\scriptsize OR} is realized by the gadget
$G_{\text{{\scriptsize OR}}}$ with vertices $\{1,2,3,4\}$ and edges
$\{\{1,3\},\{1,4\},\{2,3\},\{2,4\}\}$, where $1$ and $2$ are the input
vertices. The output vertex is vertex $4$. Vertex $3$ is initially colored black:%

\begin{figure}
[h]
\begin{center}
\includegraphics[
height=0.7185in,
width=1.4981in
]%
{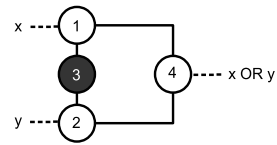}%
\caption{The gate for the function {\scriptsize OR}. }%
\label{or}%
\end{center}
\end{figure}

\emph{Proof of Claim 2. }If no action is taken then the final coloring of the
gadget is all white, but for vertex $3$. If we color vertex $1$ black then the
color-change rule implies that vertex $4$ is black at step $2$. The same holds
for vertex $2$ and for vertex $1$ and vertex $2$ together. In fact,
$\{1,3\},\{2,3\},\{1,2,3\}$ are zero forcing sets for
$G_{\text{{\scriptsize OR}}}$, able to propagate the color for inducing the
next step of the computation.

It is important to observe that zero forcing does not realize the function
{\scriptsize NOT}, since when a vertex is colored black, it can not change
color anymore. The consequence is that zero forcing does not realize universal
computation (any Boolean function can be implemented using {\scriptsize AND},
{\scriptsize OR} and {\scriptsize NOT} gates) but monotone Boolean functions
only. This concludes the proof.
\end{proof}

\bigskip

It may be worth observing the following points:

\begin{itemize}
\item Notice that extra vertices forming \emph{delay lines} may be needed to
assemble a circuit such that the output produced by zero forcing in parallel
gates is syncronous. However, given our choice of gadgets, exactly $2$ time
steps are required for output of zero forcing in $G_{\text{{\scriptsize AND}}%
}$ and $G_{\text{{\scriptsize OR}}}$. At time step $3$ the color-change rule
acts on the next gate in the circuit. There is then a convenient distinction
between internal and external time:\ \emph{internal time} refers to the zero
forcing steps inside the gadgets/gates; \emph{external time} refers to the
time steps of the computation.

\item The gadgets $G_{\text{{\scriptsize AND}}}$ and
$G_{\text{{\scriptsize OR}}}$ have three and four vertices, respectively. By
inspection on all possible combinations of white and black vertices for graphs
with at most four vertices, we can observe that we have chosen the smallest
possible gadgets, in terms of number of vertices and edges, realizing the two
functions. One might think that the gate {\scriptsize OR} is realized also by
the gagdet with three vertices in the figure:%

\begin{figure}
[h]
\begin{center}
\includegraphics[
height=0.7185in,
width=1.4981in
]%
{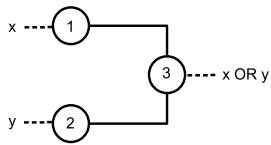}%
\caption{A gate for the function {\scriptsize OR}, where color-change rule
does not move the input forward. }%
\end{center}
\end{figure}

Although the gadget implements the {\scriptsize OR} correctly, it cannot be
used as an initial or intermediate gate of a circuit, since in this gadget the
color-change rule does not move fowards the output to the next gate, but it
halts at vertex $3$:%

\begin{figure}
[h]
\begin{center}
\includegraphics[
height=0.7185in,
width=1.2152in
]%
{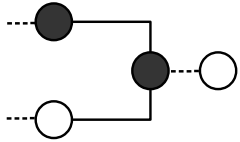}%
\caption{The figure shows that an {\scriptsize OR }gate in which all vertices
are initially white does not move the input forward.}%
\end{center}
\end{figure}

\item Let us consider the gadget $G_{\text{{\scriptsize OR}}}$. If we color
vertex $1$ black then the color-change rule implies that vertex $4$ is black
at step $2$. Suppose that vertex $2$ is colored white at step $1$. At step $2$
the gate has computed the {\scriptsize OR} function in vertex $4$ with input
$\{0,1\}$. At step $2$ vertex $2$ is also colored black under the action of
the color-change rule, because this is the unique white neighbour of vertex
$3$. This is necessary in order for the computation to proceed using the
output (black vertex $4$). So, for all inputs with output $1$, the vertices of
$G_{\text{{\scriptsize OR}}}$ are black after two steps of the internal time.
Such behaviour is discussed in more detail in the next section.

\item It is straightforward to realize the operation {\scriptsize COPY}:%

\begin{figure}
[h]
\begin{center}
\includegraphics[
height=0.7202in,
width=1.1149in
]%
{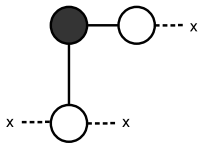}%
\caption{The gate for the function {\scriptsize COPY}. }%
\label{figucopy}%
\end{center}
\end{figure}

\end{itemize}

\section{Back forcing}

If each Boolean variable in the input of a circuit is set to $1$, then the
vertices of the circuit that are initially colored black form a zero forcing
set. However, this is not the only situation in which we have a zero forcing
set. The next figure gives an example:%

\begin{figure}
[h]
\begin{center}
\includegraphics[
height=1.5221in,
width=2.009in
]%
{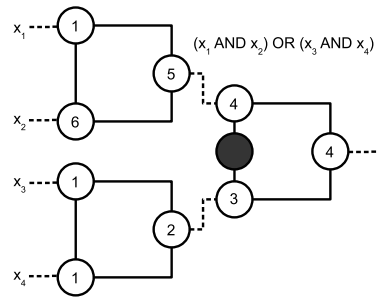}%
\caption{A circuit computing the Boolean function $(x_{1}$ {\scriptsize AND}
$x_{2})$ {\scriptsize OR} $(x_{3}$ {\scriptsize AND} $x_{4})$. The circuit
exhibits the phenomenon of back forcing.}%
\end{center}
\end{figure}

This is a circuit computing the Boolean function $(x_{1}$ {\scriptsize AND}
$x_{2})$ {\scriptsize OR} $(x_{3}$ {\scriptsize AND} $x_{4})$. The number in
the vertices of the figure specify the internal time step at which the vertex
is black; the vertices labeled by $1$ are initially colored black. The output
of the circuit is $1$ at step $4$ and at step $6$ of the internal time the
vertices encoding the input of the function are all colored black. This can
happen if and only if three of the input vertices are colored white at
internal time $1$.

The phenomenon will be called \emph{back forcing}, because it is induced by
the color-change rule acting backwards with respect to the direction from
input to output in the whole circuit. The gadget $G_{\text{{\scriptsize AND}}%
}$ exhibits back forcing conditionally on having input $\{0,1\}$. The type of
back forcing in $G_{\text{{\scriptsize AND}}}$ can be called \emph{transmittal
back forcing}, because if something back forces its output black then the gate
transmits the back force, \emph{i.e.}, it modifies the color of the output
vertex in a gate used previously. The figure clarifies the dynamics:%

\begin{figure}
[h]
\begin{center}
\includegraphics[
height=1.5957in,
width=2.8605in
]%
{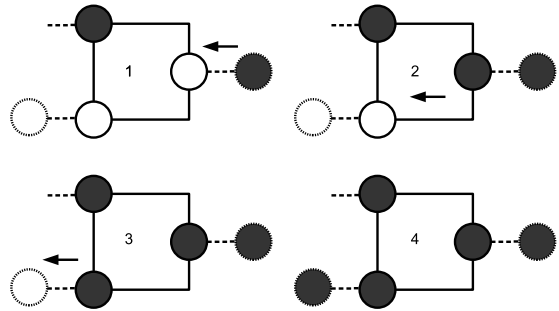}%
\caption{The steps of back forcing.}%
\end{center}
\end{figure}

The gadget $G_{\text{{\scriptsize OR}}}$ needs to force an input forward in
order to color black one of the output vertices adjacent to its inputs and in
another gate. In this sense, $G_{\text{{\scriptsize OR}}}$ does not have
transmittal back forcing. In other words, a gate at external time $t$, can not
back force its color into $G_{\text{{\scriptsize OR}}}$ at external time
$t+1$. In contrast, the circuit $(x_{1}$ {\scriptsize AND} $x_{2})$
{\scriptsize OR} $(x_{3}$ {\scriptsize AND} $x_{4})$ can initiate back forcing
as described above (when it an intermediate element in the circuit).

We can also slow down back forcing, by including appropriate \emph{delay
lines} -- for example, by adding extra vertices in each gadget or between
them. Alternatively, we could consider delay lines directly embedded in the
structure of the gadgets implementing the logical gates.

Also, back forcing can be avoided completely by including the gadget below.
The gadget acts as a \emph{filter}. In some sense, the filter can be
understood as an \emph{electronic diode} allowing zero forcing only in one direction:%

\begin{figure}
[h]
\begin{center}
\includegraphics[
height=0.7193in,
width=1.5655in
]%
{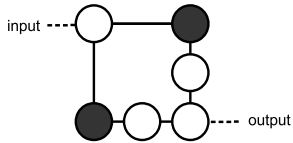}%
\caption{A gadget acting as a filter: its role is to avoid back forcing. }%
\end{center}
\end{figure}

In relation to the circuit for the function $(x_{1}$ {\scriptsize AND}
$x_{2})$ {\scriptsize OR} $(x_{3}$ {\scriptsize AND} $x_{4})$, it may be
interesting to see that if there are two parties each one chosing the input of
one of the two {\scriptsize AND} gates, and each one having access to only the
corrisponding vertices, given the back forcing, the parties can then learn the
output of the circuit by looking at the color of their vertices at the end of
the computation, except when a party chooses $(0,0)$ (\emph{i.e.}, white, white).

\section{Universality}


Despite the fact that the color-change rule induces a non-reversible process
(black coloring cannot be undone) a simple modification of the encoding
strategy allows us to implement universal, and hence also reversible, computation.

\begin{figure}
[h]
\begin{center}
\includegraphics[
height=0.8426in,
width=1.6968in
]%
{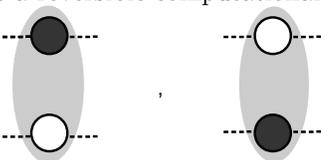}%
\caption{Physical bits for $0$ and $1$ in a dual rail encoding. }%
\label{figudual}%
\end{center}
\end{figure}

\begin{figure}
[h]
\begin{center}
\includegraphics[
height=1.0191in,
width=1.5646in
]%
{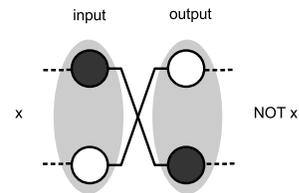}%
\caption{In a dual rail encoding the logical {\scriptsize NOT }can be
implemented by swapping the physical bits. }%
\label{figunot}%
\end{center}
\end{figure}

The idea is to adopt a \emph{dual rail} strategy, where two vertices are
employed to encode a single \emph{logical bit}. Specifically, as shown in
Fig.~\ref{figudual}, in this scheme we associate the logical bit 0 to a
configuration in which (say) the first vertex is colored in black while the
second is kept white, and the logical 1 to the opposite configuration (i.e.
the first vertex being left white and the second one being colored black).
With such encoding we can now design the gate {\scriptsize NOT} by simply
drawing a graph in which the nodes are exchanged at the output (see
Fig.~\ref{figunot}). Also a dual rail {\scriptsize AND} gate can be easily
realized. Universal computation is hence achieved by constructing a
{\scriptsize NAND} gate via concatenation of {\scriptsize AND} with
{\scriptsize NOT} and by observing that the {\scriptsize COPY} gate for the
dual rail encoding is simply obtained by just applying to both the nodes that
form a bit the transformation of Fig.~\ref{figucopy}. Once universal
computation has been achieved, we can easily turn it into a reversible one,
\emph{e.g.}, by building a Toffoli gate~\cite{toffoli}. This to remark that
even if zero forcing is an irreversible process, it can still be used to
induce a reversible computational dynamics. %

%

\section{Conclusions}

We have shown that all monotone Boolean functions can be realized by zero
forcing in a graph constructed by \emph{gluing} together the copies of two
types of subgraphs/gadgets corresponding to the Boolean gates
{\scriptsize AND} and {\scriptsize OR}. We have briefly discussed the
minimality of such gadgets in terms of vertices and edges.

We have highlighted a back forcing action. Back forcing has an effect on the
coloring of gates already used, as a function of what has happened in the
\textquotedblleft future\textquotedblright, \emph{i.e.}, at a later stage of
the computation. Because of the relation between zero forcing and minimum
ranks, the model described here is amenable to be studied with linear
algebraic tools, potentially suggesting a novel direction in the analysis of
monotone Boolean functions.

Finally, we have shown that universal computations can be obtained by zero
forcing by simply adopting a dual rail encoding.

An open problem suggested by the paper is to understand the link between zero
forcing and the dynamics at the basis of other unconventional models of
computation, like, for example, the billiard ball computer -- introduced as a
model of reversible computing \cite{ft} --, models involving geometric
objects, and dominos \cite{de}.

\bigskip

\begin{acknowledgments}
This work has been done while DB was with the Blackett Laboratory at Imperial
College London, supported by EPSRC grant EP/F043678/1. VG acknowledges support
by the FIRB-IDEAS project (RBID08B3FM). SS is a Newton International Fellow.
\end{acknowledgments}

\end{document}